# Association between quality of clinical practice guidelines and citations given to their references

Jens Peter Andersen[1,2]

[1] *jepea@rn.dk*
Aalborg University Hospital, Medical Library, DK-9000 Aalborg (Denmark)

[2] *jpa@iva.dk*
Royal School of Library and Information Science, DK-9220 Aalborg E (Denmark)

**Abstract**
It has been suggested that bibliometric analysis of different document types may reveal new aspects of research performance. In medical research a number of study types play different roles in the research process and it has been shown, that the evidence-level of study types is associated with varying citation rates. This study focuses on clinical practice guidelines, which are supposed to gather the highest evidence on a given topic to give the best possible recommendation for practitioners.
The quality of clinical practice guidelines, measured using the AGREE score, is compared to the citations given to the references used in these guidelines, as it is hypothesised, that better guidelines are based on higher cited references.
AGREE scores are gathered from reviews of clinical practice guidelines on a number of diseases and treatments. Their references are collected from Web of Science and citation counts are normalised using the item-oriented z-score and the $PP_{top-10\%}$ indicators.
A positive correlation between both citation indicators and the AGREE score of clinical practice guidelines is found. Some potential confounding factors are identified. While confounding cannot be excluded, results indicate low likelihood for the identified confounders. The results provide a new perspective to and application of citation analysis.

**Conference Topic**
Old and New Data Sources for Scientometric Studies: Coverage, Accuracy and Reliability (Topic 2) and Science Policy and Research Evaluation: Quantitative and Qualitative Approaches (Topic 3)

**Introduction**
While most scientometric studies of research publications focus on standard journal articles, it has been suggested several times that other document types may play a role in research assessment as well (Lewison, 2002, 2003; van Leeuwen, Costas, Calero-Medina, & Visser, 2012). Lewison (2002, 2003) in particular has emphasised the possibilities of various document types in the medical fields. One document type particular to that field is the clinical practice guideline. The purpose of these guidelines is to gather the best evidence of the treatment of diseases to ensure the best possible treatment at hospitals, clinics and general practices (The AGREE Collaboration, 2003). One might therefore expect these guidelines to build on the highest quality research available.

Studies have shown a connection between citation scores and the evidence-level of clinical study types (e.g. Andersen & Schneider, 2011; Kjaergard & Gluud, 2002; Patsopoulos, Analatos, & Ioannidis, 2005). These study types are hierarchically ordered, assigning greater importance to the evidence found in high-level studies, such as meta-analyses and randomised, controlled trials (RCT), than lower level studies, such as case studies (Greenhalgh, 2010). This hierarchy is widely applied in different areas of health research, and the connection between citations and evidence levels indicates that high-evidence studies are indeed, on average, used more than other. We might thus speculate if not references used in clinical practice guidelines are cited more on average than other papers, if the guidelines indeed represent the best available evidence on a topic.

Several studies have indicated that clinical practice guidelines are created very differently, with great variation in scope, rigor, clinical recommendation and overall quality (e.g. Burda, Norris, Holmer, Ogden, & Smith, 2011; Ferket et al., 2010, 2011; Freel et al., 2008; Gallardo et al., 2010; Kis et al., 2010). Many of these reviews of clinical practice guidelines use the AGREE[1] instrument (The AGREE Collaboration, 2003) to assess six major aspects of guideline development. Especially one aspect, *rigor of development*, is directly related to the evidence found in the background literature.

The hypothesis presented here is that there is a positive correlation between the rigor of development AGREE score of clinical practice guidelines and the citations given to the literature references in these clinical practice guidelines. If this correlation can be observed it points at an association between clinical evidence, citations and the development of clinical practice guidelines. The association cannot be assumed to be causative, however, but would still provide valuable insights into the inclusion and interpretation of a new document type in research assessment.

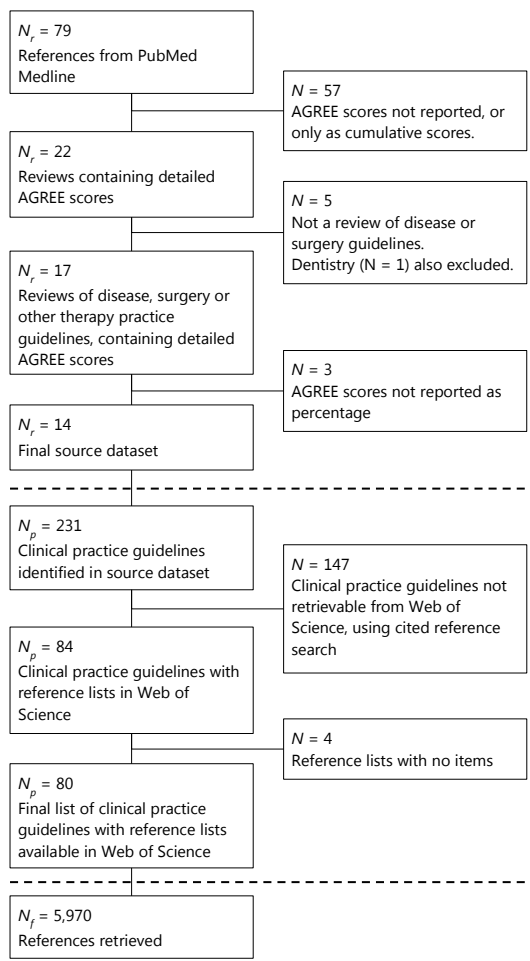

**Figure 1 - Flowchart of inclusion and exclusion criteria**

*Paper outline*

The following section outlines the acquisition of data, from reviews of clinical guidelines (top-level) to the actual guidelines and their references. This is followed by a presentation of the citation indicators used to assess the citation impact of the references of the guidelines. These references will also be discussed further in the results section which begins with an analysis of the citations given to the references, to test if they are representative of a standard citation distribution. The section concludes with a correlation analysis of the tested indicators versus the AGREE scores. All results are discussed in the final section and known weaknesses of this study are presented and discussed with respect to the findings.

**Materials and Methods**

The AGREE instrument consists of 23 key items organised in six domains, with the intention of describing various aspects of guideline development. The six domains are scope and purpose; stakeholder involvement; rigor of development; clarity and presentation; applicability; and editorial independence (The AGREE Collaboration, 2003). These domain scores have been used in a number of reviews of guideline quality as a means of assessment. Each item in the six domains is rated by one or more reviewers on a 4-point scale where 1 = strongly disagree, 2 = disagree, 3 = agree and 4 = strongly agree. In some of the included studies (see below) the overall domain scores were not calculated before all reviewers agreed on one, final item score while others used combined, standardised scores [1], resulting in a more diverse score profile, but also some inter-reviewer inconsistency. In this study the domain scores are included regardless of which procedure had been used, although it could have been preferable if all scores had been standardised. The reasoning behind this decision is elaborated in the discussion.

---
[1] Appraisal of Guidelines for Research and Evaluation

$$\frac{obtained\ score - minimum\ possible\ score}{maximum\ possible\ score - minimum\ possible\ score} \quad [1]$$

*Data collection*

To obtain the AGREE scores, reviews of clinical practice guidelines were found in PubMed Medline using the query:

*"practice guidelines as topic*"[MESH] AND "agree"[TIAB] AND "review"[PTYP] AND "2007":"2011"[PDAT] AND "English"[LANG]*

This resulted in 79 English-language reviews of clinical practice guidelines from 2007 to 2011 with the term agree occurring in the title or abstract. Not all reviews included the detailed AGREE scores, e.g. only reporting a cumulative score although this is not recommended (The AGREE Collaboration, 2003). Also reviews of non-disease topics were excluded, as were those reviews containing pure AGREE scores rather than percentages. An overview of the selection process is presented in figure 1. The final set of reviews ($N_r = 14$) were used to collect references to clinical practice guidelines ($N_p = 231$), and to gather AGREE scores for these.

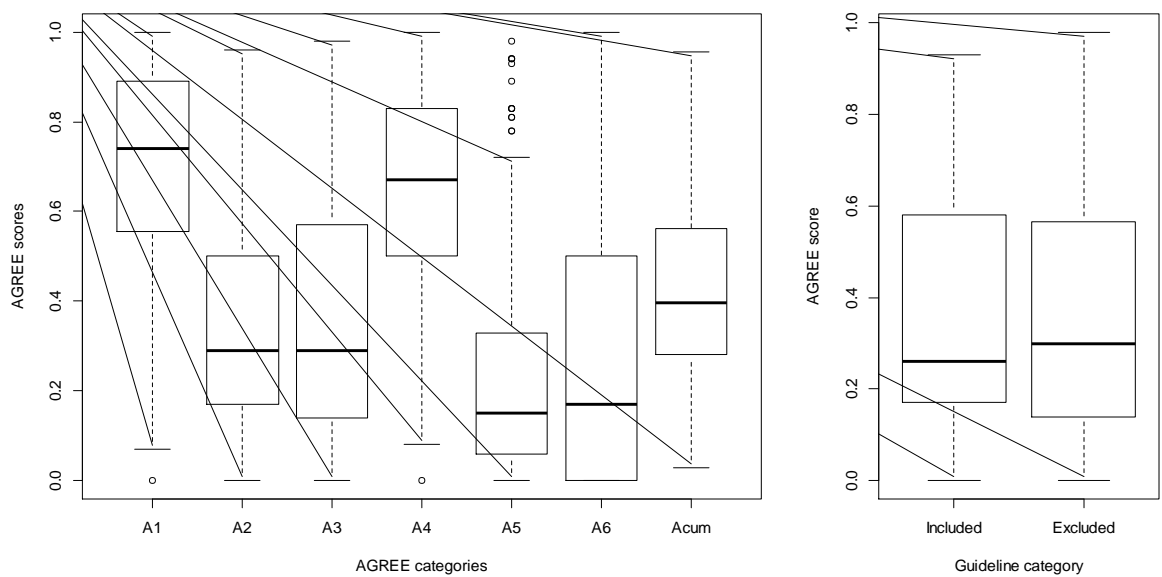

**Figure 2 – AGREE scores for all studies (left); A1: scope and purpose, A2: stakeholder involvement, A3: rigor of development, A4: clarity and presentation, A5: applicability, A6: editorial independence, Acum: cumulated, weighted scores. The figure to the right shows the difference in A3 score for guidelines included in the study and those that could not be found in Web of Science (excluded).**

As stated in the present hypotheses, it is assumed that there is a connection between the scores of the rigor of development domain (henceforth referred to as A3) and the citations given to references used by the clinical practice guidelines. It is therefore necessary to retrieve the reference lists of the above guidelines in a database where information about citations was also available. The Thomson Reuters Web of Science (WoS) was used for this purpose, and citations from all citation indices were included.

Many of the guidelines are not registered in WoS ($N = 147$), as they are often published in different channels, e.g. society websites. Those that were retrieved and contained reference lists ($N_p = 80$), however, resulted in a total of 5,970 non-unique references in their reference lists. The AGREE scores of all guidelines are shown in Figure 2, illustrating the variation between the six domains, stressing the importance of using A3 rather than a cumulated score for all six domains. The cumulated score, $A_{cum}$, in Figure 2 was calculated as a weighted average, weighting each domain by the number of items in that domain. Figure 2 also shows the difference between the A3 score for the 80 included guidelines and for the 147 excluded guidelines. As can be seen, the difference in score is very small, and the included sample can be considered representative of the complete 231 guidelines in this respect.

The references from the guidelines were retrieved from WoS, including total number of citations per January $2^{nd}$, 2013. To provide a comparison baseline, all papers published in the same journals, the same years as the included references, were also gathered from WoS, resulting in 672,819 items.

*Citation analysis*

The included clinical practice guidelines were published in a number of different areas of medicine, with varying citation potentials, and their reference lists spanned publications from 1932 to 2010. To enable comparison between these items, citation counts were normalised, using two different approaches. A somewhat direct comparison method was desired, but also an excellence-method could provide different perspectives on the meaning of citation counts as a function of agree scores. For the latter, the $PP_{top-10\%}$ indicator (Waltman et al., 2012) is considered a sensible choice, as it both focuses on the 10% most highly cited papers (a form of excellence) and remains insensitive to extremely highly cited documents. For the more direct approach, a number of normalisation procedures are available, but the item-oriented $z$-score (Lundberg, 2007) has been chosen here, as it allows normalisation of single items while also incorporating standard scores. The item-oriented $z$-score for a single item, $z_i$, is denoted as:

$$c' = \ln(c + 1)$$

$$z_i = \frac{c'_i - \bar{c}'}{S(c')}$$

, where $c_i$ is the number of citations given to a document published in a specific journal a specific year, taken from $c$, the entire distribution of citations to papers in that journal that year. $\bar{c}'$ is the average of the $c'$ distribution and $S(c')$ is the standard deviation of the $c'$ distribution. The cumulative $z$ for a clinical practice guideline is the average $z_i$ for all references in the reference list. Values of $z$ indicate the number of standard deviations the citations diverge from the mean. Positive values indicate higher than average scores.

The $PP_{top-10\%}$ indicator is simply found by comparing the citations to each paper to those of all other papers published in the same journal and year. If the citation score is placed in the highest decile, it is counted as 1, if not as 0. The $PP_{top-10\%}$ is the average of this distribution, resulting in a percentage of papers in the reference list that are considered excellent. If this percentage is higher than 10%, the references could be considered to be more excellent than standard publications. One should however be careful to draw the same conclusions in the setting of this study than one would for e.g. the publications of a university, as is the case in the Leiden ranking (Waltman et al., 2012). As all references in a reference list are de facto cited at least once, and there is an increased chance that highly cited documents are used as references, a $PP_{top-10\%}$ over 0.1 should not be interpreted as better than average. The interpretation of whether the hypothesis of this study has merit thus depends on whether an increase in $PP_{top-10\%}$ or $z$ can be observed as a function of A3.

Both citation indicators are calculated for individual references and subsequently cumulated for the guidelines. By doing so, both indicators enable comparison between guidelines, without bias from the length of reference lists, which varies greatly (from 4 to 627 with a median of 81).

All data extraction and calculations were performed using R version x64 2.15.2 (R Development Core Team, 2010)

**Results**

The citations given to references from the guidelines are plotted as a decreasing function of rank in Figure 3. The shape of the curve is as could be expected for a typical citation distribution, indicating that the sample of references is comparable to other citation distributions. The distribution of $z$-scores reveals that the citations are higher than the background population though, as the median $z$-score is 0.9, and thus almost one standard deviation higher than the expected average. This is also illustrated in Figure 4, showing the empirical cumulative density function for the $z$-score of all references.

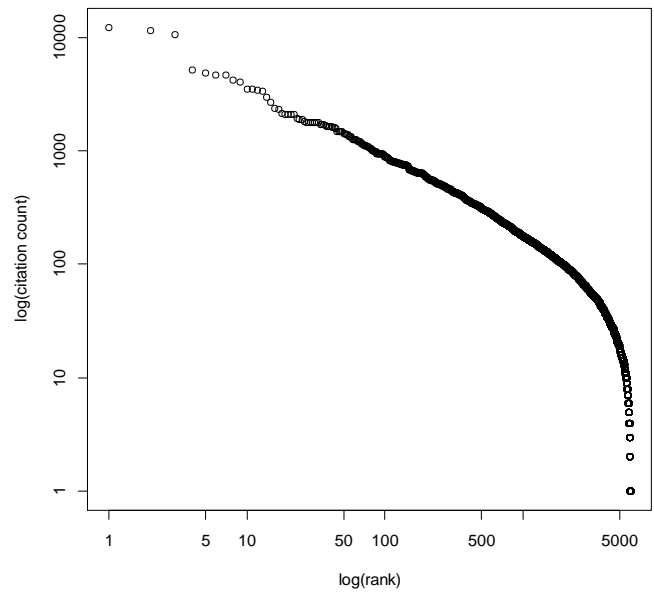

**Figure 3 – citation counts as a function of rank, double-log scales**

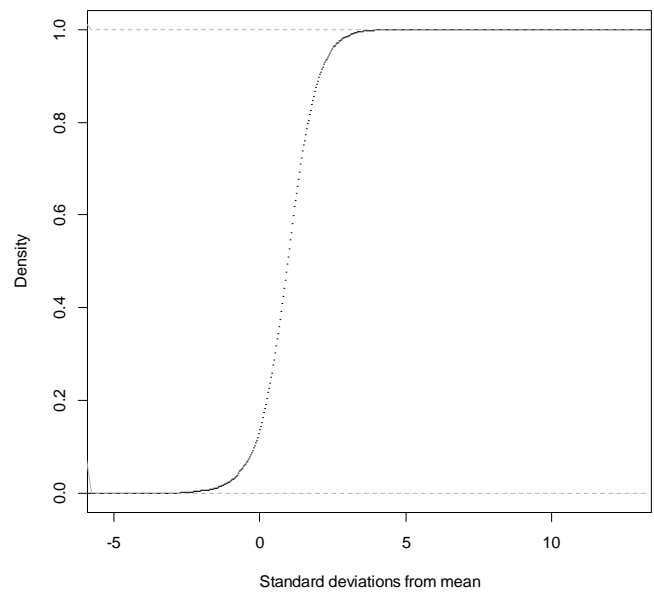

**Figure 4 – empirical cumulative density function of z-scores for references**

As is stated in the above, one might expect citation distributions for documents retrieved from reference lists to be higher than complete citation distributions of journals, as they represent documents that are actually used as references while a certain proportion of all journal articles remain uncited even after several years.

To test the correlation between A3, $z$ and PPtop-10% respectively the two citation indicators were plotted as functions of A3 in Figure 5.

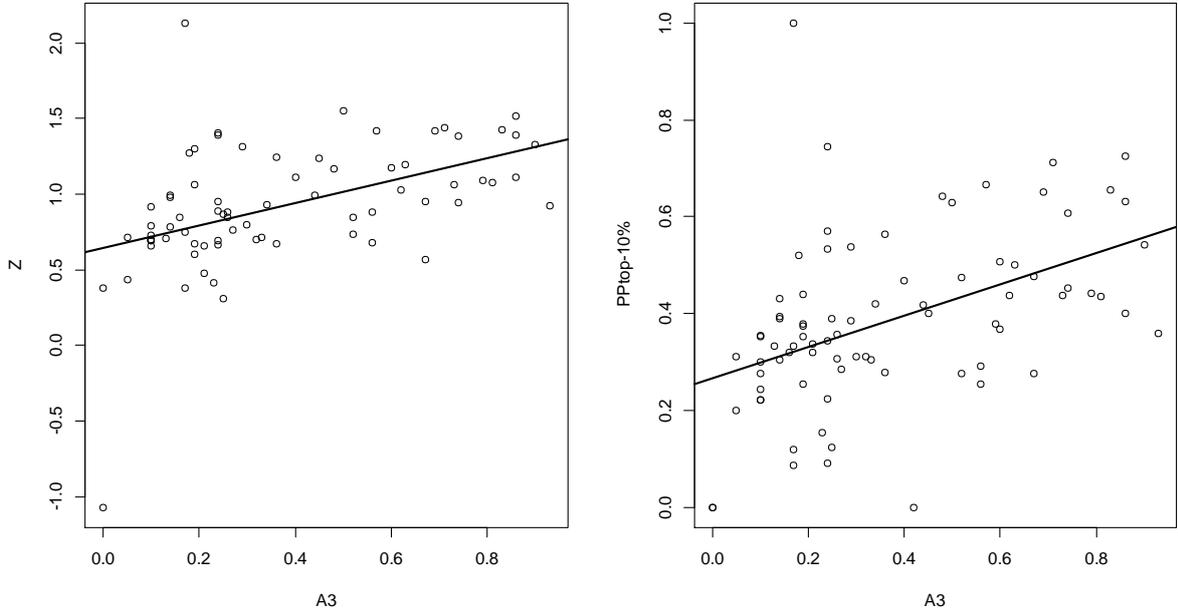

**Figure 5 – Cumulative $z$-scores of clinical practice guidelines as a function of A3 (left), and $PP_{top-10\%}$ of clinical practice guidelines as a function of A3 (right). Solid lines are fitted linear regression lines.**

As can be seen from Figure 5, there is a seemingly positive correlation for both citation indicators. Linear regression was performed on the data and regression lines were fitted to the plots (solid lines). The residuals from the regressions showed no systematic error, and Q-Q plots of the residuals showed approximately normally distributed residuals (Figure 6), with only few outliers. Thus there is no reason to assume more complex correlations, despite the large variation of data.

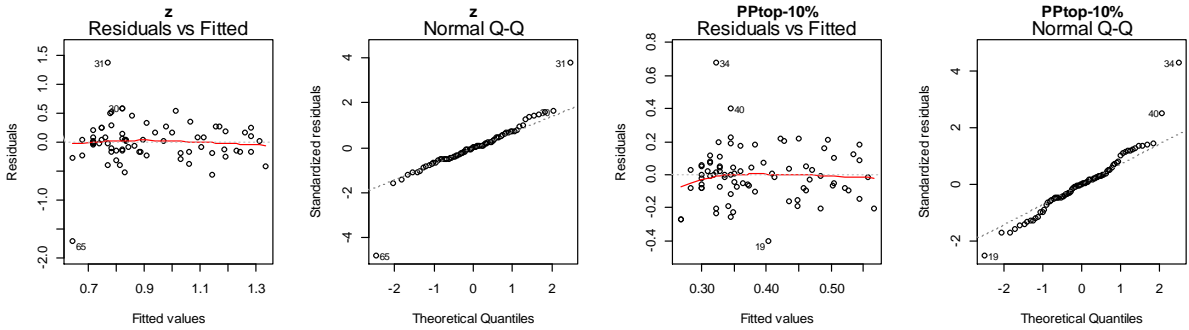

**Figure 6 - Diagnostic plots for linear regressions. Residuals and Q-Q for $z$ to the left and to the right for $PP_{top-10\%}$.**

The increase in the two indicators as a function of A3 can be considered very high. For the $z$-score, the regression line increases from .6 to 1.4 standard deviations. This accounts for a very large increase, and can be interpreted as moving from "mainstream" to "top science". For the $PP_{top-10\%}$ indicator a similar pattern emerges, with an increase from approximately .27 to .59 (59% of all references belong to the top-10% of the highest cited papers in their respective journals).

Several other factors may be influencing the correlations, some of which are not directly related to bibliometric data and some are. There is a risk of confounding error from the number of references of the clinical practice guidelines as well as the citations given to the guidelines themselves. The two citation indicators are thus plotted as functions of number of references and citations per year respectively, displayed in Figure 7. The first column in Figure 7 shows the correlation between the citation indicators and number of references in the clinical practice guidelines, the middle column the correlation between citation indicators and the annual citations received by the clinical practice guidelines. The latter is potentially related to the mechanism described by this study, therefore the AGREE scores A3 and $A_{cum}$ are plotted in the final column, as a function of annual citations received. All plots have regression lines to show the general tendencies. The statistical strength of the $PP_{top-10\%}$ regressions is low (not statistically significant, $p>0.05$), while the other regressions are somewhat stronger ($p<0.05$), especially the A3 and $A_{cum}$ regressions ($p<0.01$). This should also be apparent from a visual interpretation of the six plots in Figure 7, and the meaning of these will be discussed further in the following section.

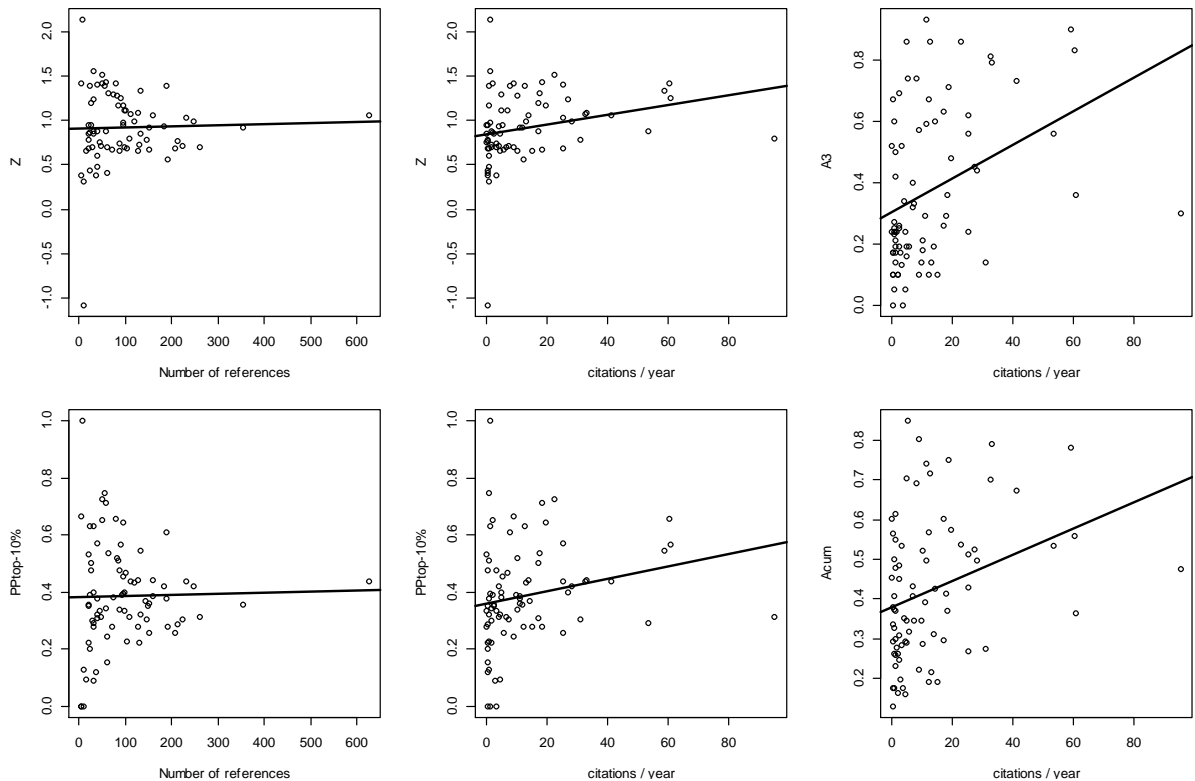

**Figure 7 - Citation indicators as functions of number of references in clinical practice guidelines (left), citations per year given to clinical practice guidelines (middle) and correlation between A3 (top-right), $A_{cum}$ (bottom-right) and citations per year.**

**Discussion**

The purpose of this study is to investigate the potential association between AGREE assessments of clinical practice guidelines and the citation scores of references used by these guidelines. The hypothesis states a positive correlation, as guidelines scoring highly on the AGREE domain *rigor of development* are assumed to build on better evidence which is here speculated to be associated with higher citation scores based on increased citations to study types associated with higher citation rates (Andersen & Schneider, 2011; Kjaergard & Gluud, 2002; Patsopoulos et al., 2005). Positive correlations are found, and the increased effect from the lowest A3 scores to the highest is considered very large. While this informs us about a strong association between citations to references and guideline development quality we cannot assume any mechanisms behind the association, as several models might explain the behaviour. From a bibliometric viewpoint, a plausible and interesting cause for the positive relationship is related to the interpretation of citations and the indicated relationship with clinical evidence. This study adds an argument to the hypothesis that clinical studies receive more citations if they provide more (relevant) evidence. An important caveat is the distinction between clinical research and other medical research, e.g. biomedical, as clinical research mostly is concerned with the testing and application of treatments. Therefore clinical practice guidelines will generally rely on clinical research, as it is usually closer to practice as more basic research areas. If biomedical studies are not included in clinical practice guidelines it is thus not a question of missing evidence, but rather of different evidence types, and conclusions about citability or citation scores should thus not be transferred from clinical research to biomedical.

A different, plausible mechanism, which is likely also affecting the results presented here, is the increased focus on the references used in clinical practice guidelines. It is not unlikely that articles will receive more attention, once they have been cited by a clinical practice guideline, ultimately leading to an increase in citations. This mechanism could be investigated further, but would require more elaborate data on citations per year than was acquired for this study.

As stated above, there is a danger of confounding errors from other sources when doing these types of analysis. Two potential confounding factors were mentioned previously, namely number of references in the guidelines and the citations given to the guidelines. The number of references might influence the results either way, as e.g. a short reference list might indicate a focus on only the very best evidence, but also a failure to include important evidence. The plots with number of references as the independent variable illustrate much larger variation for short reference lists than for the longer ones, but also that the reference list length is practically unrelated to the citation indicators, with almost horizontal regression lines roughly around the sample mean of the citation indicators. There is however a positive correlation between the annual citation counts of the clinical practice guidelines and the citation impact of their references. This effect can be explained by both of the causative models presented above. If the mention of citations in guidelines leads to higher citation scores for their references, it could be expected that the use of an article as a reference in a highly cited guideline would lead to a greater attention increase than the use in a less cited guideline. The evidence-mechanism, on the other hand, implies that the studies with the highest AGREE-scores would refer to the best evidence and by proxy contain good evidence themselves. Therefore one would also expect the citation scores of the guidelines with the highest AGREE-scores to be higher than those with low AGREE-scores. This is indeed the case, as can be seen in the rightmost plots in Figure 7, showing a clearly increasing function for both A3 and $A_{cum}$ when plotted as a function of citations per year to the clinical practice guidelines. This provides further evidence that the proposed mechanism has merit, but both causative models may yet be active at the same time.

The A3 score used in the present study is far from being a perfect measurement of the degree of foundation on evidence. The measurement is subjective and designed for an entirely different purpose as applied here, which however has the advantage of making the above usage unobtrusive, thereby removing one type of potential bias. More critical weaknesses of the measure, in the current context, are the differences in application and the meaning of the individual items of the A3 domain score. As was mentioned in the methods section, some guideline reviews implemented an interpretation of the AGREE instrument where all assessors needed to agree on each item score, where other reviews used a standardised approach allowing for inter-assessor inconsistency. One might argue for the benefits of each approach, and the arguments from a clinical standpoint are likely to be different than from a bibliometric one. While a uniform approach would have been preferable, the difference was not considered a major problem in this context, however, as the focus is on the overall increase in effect from the lowest scores to the highest. Given the broadly distributed scores in each AGREE domain, the potential error from the different approaches is considered negligible, and both approaches are included.

Not all of the seven items in the A3 domain are directly related to the background literature, although most are. It is not possible to deduce how well each guideline scored in each item from the final score, and only very few reviews reported the full background data. Two guidelines with equal scores might thus represent different scores for the items most closely related to the mechanism described in this study. For the extreme cases, i.e. the lowest and highest scores, the problem is not as large as for the mid-range, where the potential for error is much larger. The problem is thus not as relevant when regarding the overall effect, and some of the variance could be explained by this issue.

The final weakness of the study to be discussed here is the use of journal-specific normalisation methods for citation indicators. While the normalisation as such is regarded as a strength, as it allows comparison between papers with different citation potentials, it is debatable whether other normalisation methods might be more appropriate. A common normalisation procedure is the field-normalisation, in which citations are normalised with respect to the entire field rather than merely the journal. It has been argued by e.g. Leydesdorff & Bornmann (2011) that field definitions are not clearly representative of the citation potentials of the individual journals in some field category definitions. It is thus not given that a field-normalisation would necessarily improve the design of this study.

The results presented here indicate an association between clinical evidence, citation rates and the quality of clinical guidelines dependant on the degree to which they are based on evidence. This association may be useful for research assessment, studies on clinical impact and provides insight into one of the many mechanisms behind citations. The use of a study as a reference in a well-developed clinical practice guideline may be seen as a mega-citation in some respects, as it very clearly indicates usefulness in a practice-setting, which is otherwise difficult to capture with traditional citation measures. It has been stressed that impact in a broad perspective has many other aspects than research (citation) impact (Kuruvilla, Mays, Pleasant, & Walt, 2006). While many of the impact types described by Kuruvilla et al. are not related to citation analysis at all, the observations in this study allow us to broaden the application of citation analysis to e.g. health policy and practice impact studies, while in no way claiming to cover all facets of these complex subjects.

# References

Andersen, J. P., & Schneider, J. W. (2011). Influence of study design on the citation patterns of Danish, medical research. *Proceedings of the ISSI 2011 Conference* (pp. 46–53).

Burda, B. U., Norris, S. L., Holmer, H. K., Ogden, L. a, & Smith, M. E. B. (2011). Quality varies across clinical practice guidelines for mammography screening in women aged 40-49 years as assessed by AGREE and AMSTAR instruments. *Journal of clinical epidemiology*, *64*(9), 968–76. doi:10.1016/j.jclinepi.2010.12.005

Ferket, B. S., Colkesen, E. B., Visser, J. J., Spronk, S., Kraaijenhagen, R. A., Steyerberg, E. W., & Hunink, M. G. M. (2010). Systematic Review of Guidelines on Cardiovascular Risk Assessment. *Archives of internal medicine*, *170*(1), 27–40.

Ferket, B. S., Genders, T. S. S., Colkesen, E. B., Visser, J. J., Spronk, S., Steyerberg, E. W., & Hunink, M. G. M. (2011). Systematic review of guidelines on imaging of asymptomatic coronary artery disease. *Journal of the American College of Cardiology*, *57*(15), 1591–600. doi:10.1016/j.jacc.2010.10.055

Freel, A. C., Shiloach, M., Weigelt, J. a, Beilman, G. J., Mayberry, J. C., Nirula, R., Stafford, R. E., et al. (2008). American College of Surgeons Guidelines Program: a process for using existing guidelines to generate best practice recommendations for central venous access. *Journal of the American College of Surgeons*, *207*(5), 676–82. doi:10.1016/j.jamcollsurg.2008.06.340

Gallardo, C. R., Rigau, D., Irfan, A., Ferrer, A., Caylà, J. A., & Bonfi, X. (2010). Quality of tuberculosis guidelines : urgent need, *14*(October 2009), 1045–1051.

Greenhalgh, T. (2010). *How to read a paper: The basics of evidence-based medicine* (4th ed.). Oxford: BMJ Books.

Kis, E., Szegesdi, I., Dobos, E., Nagy, E., Boda, K., Kemény, L., & Horvath, a R. (2010). Quality assessment of clinical practice guidelines for adaptation in burn injury. *Burns : journal of the International Society for Burn Injuries*, *36*(5), 606–15. doi:10.1016/j.burns.2009.08.017

Kjaergard, L. L., & Gluud, C. (2002). Citation bias of hepato-biliary randomized clinical trials. *Journal of clinical epidemiology*, *55*(4), 407–10.

Kuruvilla, S., Mays, N., Pleasant, A., & Walt, G. (2006). Describing the impact of health research: a Research Impact Framework. *BMC health services research*, *6*, 134. doi:10.1186/1472-6963-6-134

Lewison, G. (2002). From biomedical research to health improvement. *Scientometrics*, *54*(2), 179–192.

Lewison, G. (2003). Beyond outputs: new measures of biomedical research impact. *Aslib Proceedings*, *55*(1/2), 32–42. doi:10.1108/00012530310462698

Leydesdorff, L., & Bornmann, L. (2011). How fractional counting of citations affects the impact factor: Normalization in terms of differences in citation potentials among fields of science. *Journal of the American Society for Information Science and Technology*, *62*(2), 217–229. doi:10.1002/asi.21450

Lundberg, J. (2007). Lifting the crown - citation z-score. *Journal of Informetrics*, *1*(2), 145–154.

Patsopoulos, N. a, Analatos, A. a, & Ioannidis, J. P. A. (2005). Relative citation impact of various study designs in the health sciences. *JAMA : the journal of the American Medical Association*, *293*(19), 2362–6. doi:10.1001/jama.293.19.2362

R Development Core Team. (2010). *R: A language and environment for statistical computing*. Vienna, Austria: R Foundation for Statistical Computing.

The AGREE Collaboration. (2003). Development and validation of an international appraisal instrument for assessing the quality of clinical practice guidelines: the AGREE project. *Quality & safety in health care*, *12*(1), 18–23. Retrieved from http://www.pubmedcentral.nih.gov/articlerender.fcgi?artid=1743672&tool=pmcentrez&rendertype=abstract

Van Leeuwen, T. N., Costas, R., Calero-Medina, C., & Visser, M. S. (2012). The role of editorial material in bibliometric research performance assessments. In É. Archambault, Y. Gingras, & V. Larivière (Eds.), *Proceedings of STI 2012 Montréal* (pp. 511–522). Montréal, Canada: 17th International Conference on Science and Technology Indicators.

Waltman, L., Calero-Medina, C., Kosten, J., Noyons, E. C. M., Tijssen, R. J. W., Van Eck, N. J., Van Leeuwen, T. N., et al. (2012). The Leiden ranking 2011/2012: Data collection, indicators, and interpretation. *Journal of the American Society for Information Science and Technology*, *63*(12), 2419–2432. doi:10.1002/asi.22708